\documentclass[aps,twocolumn,amsfonts,showpacs]{revtex4}
\begin{document}
\newcommand{\bb}{\begin{equation}}
\newcommand{\ee}{\end{equation}}
\newcommand{\eqb}{\begin{eqnarray}}
\newcommand{\eqf}{\end{eqnarray}}

\preprint{}
\title{Quantum Uncertainty in Doubly Special Relativity}
\author{J. L. Cort\'es}
\email{cortes@unizar.es}
\affiliation{Departamento de F\'{\i}sica Te\'orica, Universidad de
  Zaragoza, Zaragoza 50009, Spain} 
\author{J. Gamboa}
\email{jgamboa@lauca.usach.cl}
\affiliation{Departamento de F\'{\i}sica, Universidad de Santiago de Chile,
Casilla 307, Santiago 2, Chile}
\begin{abstract}
The modification of the quantum mechanical commutators in a
relativistic theory with an invariant length scale (DSR) is
identified. Two examples are discussed where a classical behavior is
approached in one case when the energy approaches the inverse of the
invariant length which appears as a cutoff in the energy and in the
second case when the mass is much larger than the inverse of the
invariant length.     
\end{abstract}
\maketitle
\section{Introduction}
One of the basic open problems in theoretical physics is to combine in
a consistent way the classical description of the gravitational
interaction (general relativity) with quantum mechanics (QM). The analysis
of the problems that one finds in different attempts to combine these
two theories can be used as a guiding principle to the search of a
fundamental theory of quantum gravity. 

On the other hand one has direct proposals for this theory like
string(M)-theory or loop quantum gravity. But we are not able,  
presently to establish if any of these or another future proposal is the
correct theory. An alternative is to try to identify basic ingredients
that this theory should contain and whose details could be a criteria
in the future to select among different alternatives. One idea --that
has been discussed very often in this context \cite{length}-- is the possible
appearance of a fundamental length, a scale associated to gravity not
just as a dynamical scale but at the kinematical level.

Several arguments going from theoretical analysis in string theory
\cite{string} to more or less sophisticated gedanken 
experiments to measure lengths \cite{length},  lead to the conclusion that there is a
new contribution to the quantum uncertainty of gravitational origin
leading to a length scale as a minimal uncertainty in the determination
of space-time coordinates. Some attempts to identify a modification of
the quantum mechanical commutators \cite{modcom} which reproduce the
generalized uncertainty principle,  have been considered as a way to
find one of the ingredients of the quantum theory of gravity.

Another candidate for a signal of quantum gravity effects is the
modification of Lorentz symmetry at very high energies, an idea
explored intensively in the last years both from a theoretical as well
as from a phenomenological perspective \cite{lorentz}.   

Alternatively,  the limitation due to gravity to explore beyond a
minimal length, has motivated to consider the possibility of a
generalization of the relativity principle compatible with an
invariant length scale which is called doubly special relativity (DSR)
\cite{dsr1,dsr2} as a possible ingredient of the flat space-time limit of the quantum
theory of gravity. It is remarkable that one can have a generalized
relativity principle which is continuously connected with standard
Einstein special relativity (SR), compatible with all the tests
of SR and leading to corrections accessible to present or
near future experiments or even already observed in the high energy
tail of the cosmic ray spectrum \cite{dsrph1,dsrph2}. Different
proposals for DSR theories 
either based on a deformation of Poincare algebra or by considering
directly a modification of the boosts that connect inertial frames has
been considered recently from different points of view \cite{dsr1,dsr2,dsr}.   

It is still not clear how to introduce space-time in DSR \cite{spacetime}. The aim of
this paper is to study the modification of QM at the level of
commutation relations associated to a modified relativity principle
based on a nonlinear realization of Lorentz transformations in
energy-momentum space (which is a common ingredient of the different
proposals of DSR theories) together with a simple prescription for the
space-time sector.  
We conjecture that the modifications of QM obtained in this way are a
remnant of the fundamental theory of QG in the flat space-time limit. 

\section{Modified Quantum Mechanical Commutators}
A nonlinear realization of Lorentz transformations in energy-momentum
($E$, ${\bf p}$) space parametrized by an invariant length $\ell$ can
be defined \cite{MS2} by the relations
\begin{equation}
\epsilon = E \; f\left(\ell E, \ell^2 {\bf p}^2\right) 
\label{ava} 
\end{equation}
\begin{equation}
\pi_i = p_i \; g\left(\ell E, \ell^2 {\bf p}^2\right)
\label{avb}
\end{equation} 
where ($\epsilon$, ${\bf \pi}$) are auxiliary linearly transforming
variables which define the nonlinear Lorentz transformation of the
physical energy-momentum ($E$, ${\bf p}$). Then we have two functions
of two variables ($f$, $g$) which parametrize the more general
nonlinear realization of Lorentz transformations, with rotations
realized linearly, depending on a dimensional scale. The condition to
recover the special relativistic theory in the low energy limit
reduces to the condition $f(0, 0) = g(0, 0) = 1$. Each choice of
the two functions $f$, $g$ will lead to a generalization of the
relativity principle with an invariant length scale $\ell$. Lorentz
transformation laws connecting the energy-momentum of a particle in
different inertial frames differ from the standard special relativistic
linear transformation laws which are recovered when $\ell E \ll 1$, 
$\ell^2 {\bf p}^2 \ll 1$. 

In order to have a quantum theory with such
a deformed relativity principle, one should find the appropriate
deformation of relativistic quantum field theory (QFT). First attempts in
this direction, based on the possible connection between a
generalization of the relativity principle and a non-commutativity of
space-time, suggesting the formulation of QFT in a non-commutative
space ($\kappa$-Minkowsky) as the appropriate deformation of QFT have
been explored \cite{drqft}. But there are general arguments that there will be
difficulties to find a realization of a deformed relativity principle
along these lines in the multiparticle sector \cite{amel1,dsrph1}. Due
to these problems we consider in 
this work a less ambitious program trying to give an implementation of
DSR at the level of quantum mechanics. The simplest way to do this is
to introduce space-time coordinates as the generators of translations
in the auxiliary linearly transforming energy-momentum variables   
($\epsilon$, ${\bf \pi}$) which then reduce to the usual space-time
coordinates of special relativity in the limit $\ell \to 0$. In this
case one does not have any signal of the modified relativity principle
at the level of the space-time coordinate commutators which are still
trivial but all the modifications appear at the level of the phase
space commutators which will be 
 \begin{equation}
\left[t, E\right] = i \hbar \frac{\partial E}{\partial\epsilon}
\label{a1a}
\end{equation}    
 \begin{equation}
\left[t, p_i\right] = i \hbar \frac{\partial p_i}{\partial\epsilon}     
\label{a1b}
\end{equation}    
 \begin{equation}
\left[x_i, E\right] = i \hbar \frac{\partial E}{\partial\pi_i}
\label{a1c}
\end{equation}    
 \begin{equation}
\left[x_i, p_j\right] = i \hbar \frac{\partial p_i}{\partial\pi_j} \,.
\label{a1d}
\end{equation}    
By considering the derivatives with respect to the auxiliary
energy-momentum variables of the equations (\ref{ava}-\ref{avb}) defining the
nonlinear realization of Lorentz transformations, one has a linear
system of equations for the partial derivatives required to calculate
the phase space commutators (\ref{a1a}-\ref{a1d}). A straightforward algebra
leads to the modified quantum mechanical commutators
\begin{equation}
\left[t, E\right] = i \hbar \frac{g + 2 \ell^2 {\bf p}^2
  \partial_2 g}{D}
\label{mca}  
\end{equation}
\begin{equation}
\left[t, p_i\right] = - i \hbar  \frac{\ell p_i \partial_1 g}
{D}  
\label{mcb}
\end{equation}
\begin{equation}
\left[x_i, E\right] = - i \hbar \ell p_i \frac{2 \ell E
  \partial_2 f}{D}
\label{mcc}  
\end{equation} 
\begin{equation}
\left[x_i, p_j\right] = \frac{i \hbar}{g} \left[\delta_{ij} - 2
  \ell^2 p_i p_j \frac{N}{D}\right] 
\label{mcd}
\end{equation} 
where
\begin{equation}
D=\left[f+\ell E\partial_1 f\right]\left[g+2\ell^2{\bf p}^2\partial_2 g\right]
-2\ell^2{\bf p}^2\partial_1 g\ell E\partial_2 f \;,
\end{equation}
\begin{equation}
N=f\partial_2 g +\ell E\left(\partial_1 f\partial_2 g -\partial_2
f\partial_1 g\right) \;.
\end{equation}
This result can be seen as an
explicit realization of a general idea that in the presence of quantum
gravity the quantum mechanical uncertainty principle, and then the
phase space commutators on which it is based, should be modified by
terms depending on an invariant length \cite{modcom}. Instead of guessing
the general structure of the generalized uncertainty principle and the
modification of the phase space commutators leading to such a
generalization of the uncertainty principle, a modification of the
commutators (\ref{mca}-\ref{mcd}) is obtained directly from a non
linear realization of Lorentz transformations in momentum space
parametrized by the functions $f$, $g$. In order to discuss the
consequences of the modifications of the quantum mechanical
commutators one has to specify the non-linear transformations of
energy-momentum. We discuss some cases in next two sections.        

\section{An example with an energy cutoff: DSR2}
When the two functions $f$, $g$ parametrizing the non linear Lorentz
transformations are independent of the momenta (i.e., when $\partial_2
f = \partial_2 g = 0$) the modified quantum mechanical commutators in
(\ref{mca}-\ref{mcd}) take a much simpler form
\begin{equation}
\left[t, E\right] = i \hbar \frac{1}{f + \ell E \partial_1 f}
\label{mc21a} 
\end{equation}
\begin{equation}
\left[t, p_i\right] = - i \hbar \ell p_i \frac{\partial_1 g}{g \left[{f + \ell
      E \partial_1 f}\right]}
\label{mc21b}  
\end{equation}
\begin{equation}
\left[x_i, E\right] = 0
\label{mc21c}  
\end{equation}
\begin{equation}
\left[x_i, p_j\right] = i \hbar \delta_{ij} \frac{1}{g} \;.
\label{mc21d}
\end{equation}  
A very simple choice for the functions $f$, $g$
\begin{equation}
f = g = \left(1 - \ell E\right)^{-1}
\end{equation}
is what is known as DSR2 \cite{dsr2} and corresponds to the simplest realization
of DSR with an energy cutoff ($E<1/\ell$) identified as the inverse of
the invariant length. The combinations of derivatives which appear in
the phase space commutators are given by
\begin{equation}
\partial_1 g = f + \ell E \partial_1 f = \left(1 - \ell E\right)^{-2}
\end{equation}
and then one has 
\begin{equation}
\left[t, E\right] = i \hbar \left(1 - \ell E\right)^2  
\label{mc22a}
\end{equation}
\begin{equation}
\left[t, p_i\right] = - i \hbar \ell p_i \left(1 - \ell E\right)
 \label{mc22b}
\end{equation}
\begin{equation}
\left[x_i, E\right] = 0  
\label{mc22c}
\end{equation}
\begin{equation}
\left[x_i, p_j\right] = i \hbar \delta_{ij} \left(1 - \ell E\right)
\;,
\label{mc22d}
\end{equation}
a result already anticipated in (\cite{MS2}) where it was obtained by
considering a possible realization of space-time coordinates as
differential operators in momentum space. The conclusion that one gets
from (\ref{mc22a}-\ref{mc22d}) is that there is a modification of the quantum
mechanical commutators which becomes relevant when the energy approaches
its maximum value and in the limit one gets a classical phase
space. This is a result that one could have anticipated from the
consistency of the quantum mechanical uncertainty principle with the
possibility to explore an arbitrarily small region 
in space-time while having a cutoff on the available energies. 
 
\section{An example with a momentum cutoff: DSR1}
Another example of a non linear realization of Lorentz transformations
corresponds to the choice of functions
\begin{equation}
f = \frac{1}{2} \left[\left(1 + \ell^2 {\bf p}^2\right) \frac{e^{\ell
      E}}{\ell E} - \frac{e^{- \ell E}}{\ell E}\right]  
\label{fdsr1}
\end{equation}
\begin{equation}
g = e^{\ell E}
\label{gdsr1}
\end{equation}
The relation between the energy and momentum for a particle of mass
$m$ is given by 
\begin{equation}
\left(1 - \ell^2 {\bf p}^2\right) e^{\ell E} + e^{- \ell E} =  
e^{\ell m} + e^{- \ell m}
\label{dr1}
\end{equation}
which is the dispersion relation of the model referred to as DSR1 \cite{dsr1}. One
finds 
\begin{equation}
e^{\ell E} = \frac{\cosh(\ell m) + \sqrt{\cosh^2(\ell m) - (1 - \ell^2
    {\bf p}^2)}}{1 - \ell^2 {\bf p}^2}
\end{equation}
for the energy as a function of momentum. One has in this case an
upper bound on the momentum (${\bf p}^2 < 1/\ell^2$) instead of the
energy. If one replaces the function g and its derivatives
\begin{equation}
\partial_1 g = g = e^{\ell E} {\hskip 2cm} \partial_2 g = 0
\end{equation}
into the general expression (\ref{mca}-\ref{mcd}) for the modified quantum
mechanical commutators one has 
 \begin{equation}
\left[t, E\right] = i \hbar \frac{1}{D_1}
\label{mc11a}  
\end{equation}
\begin{equation}
\left[t, p_i\right] = - i \hbar \ell p_i \frac{1}{D_1}
\label{mc11b}  
\end{equation}
\begin{equation}
\left[x_i, E\right] = - i \hbar \ell p_i \frac{2 e^{-\ell E}\ell E \partial_2 f}{D_1}  
\label{mc11c}
\end{equation}
\begin{equation} 
\left[x_i, p_j\right] = i \hbar e^{-\ell E} \left[\delta_{ij} + \ell^2
    p_i p_j \frac{2 \ell E \partial_2 f}{D_1}\right]
\label{mc11d}
\end{equation}
where
\begin{equation}
D_1= f + \ell E \left[\partial_1 f - 2 \ell^2 {\bf p}^2 \partial_2 f\right]
\;.
\end{equation}
For the particular choice for $f$ in (\ref{fdsr1})  
\begin{equation}
2 \ell E \partial_2 f = e^{\ell E}
\end{equation}
and
\begin{equation}
D_1 = \frac{1}{2} \left[\left(1 - \ell^2 {\bf p}^2\right) e^{\ell E} +
  e^{- \ell E}\right] \;.
\label{f1}
\end{equation} 
If one uses the relation between energy and momentum (\ref{dr1}), the
right hand side in (\ref{f1}) reduces to $\cosh(\ell m)$ and the phase
space commutators become
\begin{equation}
\left[t, E\right] = \frac{i \hbar}{\cosh(\ell m)}  
\label{mc12}
\end{equation}
\begin{equation}
\left[t, p_i\right] = - \ell p_i \frac{i \hbar}{\cosh(\ell m)}
 \label{mc12b}
\end{equation}
\begin{equation}
\left[x_i, E\right] =  - \ell p_i \frac{i \hbar}{\cosh(\ell m)}
 \label{mc12c}
\end{equation}
\begin{equation}
\left[x_i, p_j\right]  =  i \hbar \left[e^{-\ell E} \delta_{ij} + \ell^2 p_i p_j
\frac{1}{\cosh(\ell m)}\right] 
\label{mc12d}
\end{equation} 
We see from these expressions that when the mass $m$ is much larger
than the inverse of the length scale $\ell$ all the commutators are
(exponentially) small and a classical phase space is approached. This
result suggests the possibility to relate the transition from the
quantum behaviour at the microscopic level to the classical behavior
at the macroscopic level with the modification of quantum mechanics
induced by a modification of the relativity principle.

As a final remark one can consider the massless case where
\begin{equation}
e^{\ell E} = \frac{1}{1 - \ell |{\bf p}|}
\end{equation}
and the modified commutators are
\begin{equation}
\left[t, E\right] = i \hbar
\label{mc13a}  
\end{equation}
\begin{equation}
\left[t, p_i\right] = - i \hbar \ell p_i
\label{mc13b}  
\end{equation}
\begin{equation}
\left[x_i, E\right] = - i \hbar \ell p_i 
\label{mc13c} 
\end{equation}
\begin{equation}
\left[x_i, p_j\right] = i \hbar \left[\left(1 - \ell |{\bf p}|\right) \delta_{ij} + 
\ell^2 p_i p_j\right]
\label{mc13d}
\end{equation}
In contrast to the case of a cutoff in the energy, when the momentum
approaches its maximum value one has a non trivial limit for the
commutators which differs from the canonical commutation relations. 
 
\section{Conclusions}
The standard arguments leading to a minimum physical length beyond
which it is not possible to go in the presence of gravity assume that
in the flat space-time limit one has the standard QM uncertainty
principle. If there is a remnant of gravity in the flat space-time
limit as an invariant length and a modification of the QM commutators
along the lines of the examples considered in this work then these
arguments do not apply. In fact we have shown that in some cases one
could have that instead of an obstruction to a localization in a
physical system beyond a minimal length due to a modification of the
quantum uncertainty by gravitational effects one could have the
opposite situation where as a remnant of the QG theory the quantum
uncertainty is diluted and the system approaches to the classical
limit with no uncertainties in the high energy limit and/or for large
masses. 

If the modifications of QM suggested by the examples analyzed in
this work apply to the flat space-time limit of the QG theory then
our understanding of different physical systems should be
reconsidered. The qualitative description of physical systems on a
macroscopic scale, based on standard QM, could be altered in some
cases. Also the discussion of black hole evaporation could be
modified when one approaches the invariant length scale where quantum
black holes become classical. Even the quantum mechanical aspects of
the evolution of the Universe could differ from standard physics
expectations. An study of some of the consequences of a modification
of QM as well as a more systematic analysis of all possible QM
commutators corresponding to the different nonlinear representations
of Lorentz transformations in energy-momentum space and different ways
to introduce the space-time sector are presently under investigation.

Useful discussions with J. M. Carmona, A. F. Grillo, F. M\'endez and M. Plyuschay are
acknowledged. Work partially supported by the grants 1010596, 7010596
from Fondecyt-Chile, by M.AA.EE./AECI and by MCYT (Spain), grant
FPA2003-02948.  

\end{document}